\documentclass[%
reprint,
superscriptaddress,
amsmath,amssymb,
aps,
prb,
]{revtex4-2}

\usepackage{natbib}
\usepackage{graphicx}
\usepackage{dcolumn}
\usepackage{bm}
\usepackage{amsmath}
\usepackage{hyperref}

\begin{document}
	
	\preprint{APS/123-QED}
	
	\title{Topographic control of order in quasi-2D granular phase transitions}
	
	\author{J. G. Downs}
	\affiliation{School of Physics and Astronomy, University of Nottingham, Nottingham, NG7 2RD, United Kingdom}
	\author{N. D. Smith}
	\affiliation{School of Physics and Astronomy, University of Nottingham, Nottingham, NG7 2RD, United Kingdom}
	\author{K. K. Mandadapu}
	\affiliation{Department of Chemical and Biomolecular Engineering, University of California, Berkeley, CA 94720, USA}
	\affiliation{Chemical Sciences Division, Lawrence Berkeley National Laboratory, Berkeley, CA 94720, USA}
	\author{J. P. Garrahan}
	\affiliation{School of Physics and Astronomy, University of Nottingham, Nottingham, NG7 2RD, United Kingdom}
	\author{M. I. Smith}\email{mike.i.smith@nottingham.ac.uk}
	\affiliation{School of Physics and Astronomy, University of Nottingham, Nottingham, NG7 2RD, United Kingdom}

	\date{\today}
	
	\begin{abstract}
		We experimentally investigate the nature of 2D phase transitions in a quasi-2D granular fluid. Using a surface decorated with periodically spaced dimples we observe interfacial tension between coexisting liquid and crystal phases. Measurements of the orientational and translational order parameters and associated susceptibilities indicate that the surface topography alters the order of the phase transition from a two-step continuous one to a first-order liquid-crystal one. The interplay of boundary inelasticity and geometry, either order-promoting or inhibiting, controls the wetting of the granular crystal / fluid. This order induced wetting has important consequences, determining how coexisting phases separate spatially.
	\end{abstract}
	
	\maketitle
	
	Recent research into 2D materials development has highlighted the need for a fundamental understanding of order-disorder phase transitions in lower dimensions \cite{Li2021, Wang2018, Zanotti2016, Durand2019}. At the theoretical level this has proved challenging, with even the nature of the canonical two-dimensional hard disc phase transition only recently clarified \cite{Bernard2011}. 
	
	In 2D systems, the liquid-crystal transition often proceeds via an intermediate hexatic phase, which exhibits quasi-long-range orientational order like a crystal, but short-range positional order like a liquid \cite{Halperin1978}. Systems with hard discs, are now believed to undergo a continuous crystal-hexatic transition, followed by a first-order hexatic-liquid one \cite{Bernard2011, Engel2013}. Yet subtle changes in the inter-particle potential can alter this scenario, raising the question of how such theoretical understanding maps onto real world systems \cite{Kapfer2015}. Consequently, recent studies have explored how 2D phase transitions are influenced by factors such as polydispersity, inter-particle potential and shape \cite{Russo2017, Kapfer2015, Anderson2017, Li2020}. Experiments have played a key role, confirming these results and inspiring future research directions \cite{Briand2016, Thorneywork2017, Li2019}. Common to these studies is the idea that preventing 5/7 neighbor disclinations suppresses the hexatic phase leading to a first-order liquid-crystal phase transition.
	
	A number of quasi-2D granular studies have observed similarities between their melting behavior and that of hard discs \cite{Olafsen1998, Olafsen2005, Reis2006, Komatsu2015, Sun2016}.  Surprisingly, in these non-equilibrium systems, the crystal phase generally melts via the two-step continuous KTHNY scenario \cite{Nelson2002}. However, if the particles are highly inelastic, the liquid-crystal phase transition may become first-order \cite{Komatsu2015}. Particle inelasticity can also result in wetting by the crystalline phase. In equilibrium systems, the preferential wetting of a foreign component by the ordered or disordered phase can be governed by geometry rather than interactions, controlling the spatial phase separation at a first-order phase transition \cite{Katira2016}. Though this has been demonstrated in simulations of 2D lipid bilayers, such ideas should be generic for order-disorder transitions, both equilibrium and non-equilibrium \cite{Katira2016, Katira2018}.  
	
	Two-dimensional systems rarely exist in isolation from an external potential, e.g. a substrate or pinned impurities. Early 2D experiments found that the phase behavior of noble gases deposited on graphite depended on whether the underlying lattice was commensurate \cite{Birgeneau1981, Strandburg1988}. More recently colloidal experiments have demonstrated that a patterned substrate can control where 2D crystal phases nucleate \cite{Ganapathy2010, Mishra2016}. In a similar way, in-plane pinned particles located at regular lattice sites, can suppress the hexatic phase leading to a first-order liquid-crystal phase separation \cite{Qi2015}. 
	
	Here we explore the implications of periodic structures, both in and out of the particle plane, for phase transitions in a quasi-2D granular system. By controlling the topography of the surface and lateral boundaries we manipulate the nature of the underlying liquid-crystal phase transition and demonstrate spatial control of the wetting/phase separation. 
	
	Our experiment consists of a partial monolayer of spherical particles (D=4mm) on a horizontal metal plate. The plate is subjected to vertical sinusoidal vibrations with a dimensionless acceleration $\Gamma=A(2\pi f)^2/g$ that can be varied in increments of  $\Delta\Gamma=0.013$, where $A$ is the amplitude and $f$ is the frequency (50Hz). As the particles move across the surface in a quasi-2D layer they are filmed from above using a camera (Panasonic HC-X1000, 50fps). The location of each particle is measured using the Hough Circle transform (OpenCV) with a precision of ${\sim}0.1$mm. In addition to the particle area fraction, $\phi$, the acceleration represents an intensive control variable, playing a role somewhat analogous to the temperature in an equilibrium system. However, the non-equilibrium nature of our experiment introduces important differences. For example, the granular temperature (${\sim}\langle v^2\rangle$) in co-existing phases is not in general equal as the temperature would be in an equilibrium system \cite{Clewett2019}. Despite this caveat we will use the language of “heating”/“cooling” to describe increases/decreases in the acceleration for simplicity.
	
	Two Aluminum plates were prepared with different surface topographies: the first flat and the second with a triangular array of dimples (spacing $L\sim4.62$mm, see supplementary information). Our experiment uses custom 3D printed boundaries that are hexagonal in shape (apothem = 100mm). The boundaries are printed with convex dimples on the internal faces with carefully chosen spacings, to which we attach 4mm nitrile particles.  
	
	We initialize an experiment with $\phi\sim0.82$ by ‘heating’ the system to $\Gamma\sim2.6$. At this acceleration, the entire system exhibits a disordered liquid state. We then slowly cool the system at a rate of $\dot{\Gamma}=0.0013$s$^{-1}$. At $\Gamma\sim2.0$ a single crystalline domain surrounded by a disordered phase suddenly forms (see supplementary movie 1). \autoref{fig1}a shows a boundary where the placement of boundary particles is incommensurate with the observed crystal phase on the dimpled plate (``orderphobic''). 
	Although the region of crystal fluctuates it never wets the boundary, separated by a significant region of liquid phase. 
	
	\begin{figure}
		\centering
		\includegraphics{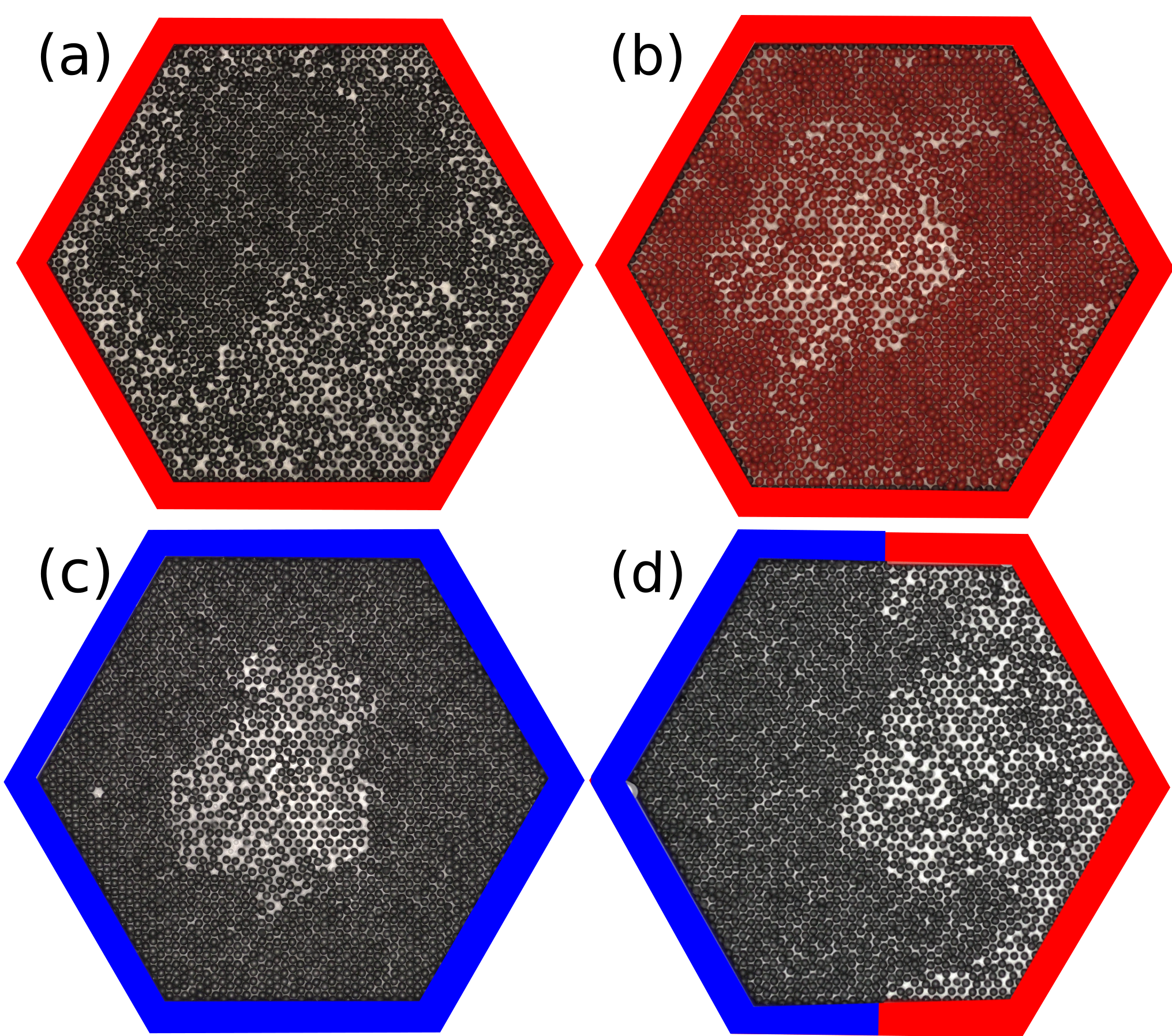}
		\caption{\textbf{Wetting of boundary by co-existing phases on a dimpled plate.} (a) Inelastic particles with an order-phobic boundary - see definition in text ($e_{pp} \sim e_{pw}$). (b) Elastic particles with an order-phobic boundary ($e_{pp} >>  e_{pw}$).  (c) Inelastic particles with an orderphilic boundary ($e_{pp} \sim e_{pw}$). (d) Inelastic particles with a 50:50 orderphobic : orderphillic boundary ($e_{pp} \sim e_{pw}$). }
		\label{fig1}
	\end{figure}
	
	The finite size of some experiments results in the edge playing a significant role in the system’s behavior \cite{Smith2017}. Yet even in larger systems the presence of in-plane structures or defects may lead to some spatially varying tendency for one or other phase \cite{Galajda2007, Qi2015, Katira2016, Reichhardt2021}. The black nitrile particles used throughout most of this study are extremely inelastic ($e\sim0.1$, see supplementary information). In \autoref{fig1}b we compare these with more elastic polypropylene particles ($e\sim0.6$), though the boundary remains the same. Upon cooling, a crystal phase with the same lattice parameters still forms. One observes competition between frustration of the crystal in the particle layers immediately adjacent to the boundaries and a tendency for the liquid phase to separate to the middle of the experiment. This occurs no matter how slowly the system is cooled, so is not a kinetically trapped configuration but a non-equilibrium steady-state.
	
	In a quasi-2D granular experiment, energy enters the system through particles interacting with the vibrating base. At large $\phi$ energy is predominantly transferred and dissipated through inter-particle ($e_{pp}$) and particle-wall ($e_{pw}$) collisions, rather than advection \cite{Smith2017}. In \autoref{fig1}b $e_{pw} << e_{pp}$, therefore the dissipative energy flux near the wall is larger than in the center of the experiment, creating a gradient in the granular temperature \cite{Smith2017}. Since the higher density crystal phase has the lower granular temperature, this therefore forms at the edge, only frustrated in the first few layers by the structure of the system boundary. In contrast, in \autoref{fig1}a the edge is preferentially wet by the liquid phase. Here, $e_{pw} \sim e_{pp}$ resulting in a more spatially uniform dissipation. This enables the boundary structure to control the location of the different phases. \autoref{fig1}c confirms this principle, using a boundary structure patterned to be commensurate with the crystalline phase (``orderphillic''). Now the crystal phase wets the boundaries, reversing the spatial phase separation (cf \autoref{fig1}a). This demonstrates that in the absence of strong gradients in dissipation, the wetting can be controlled purely by order/disorder boundary conditions. 
	
	Making use of this fact we created a hybrid boundary, where 3 sides (red) promote disorder and the remaining sides (blue) have been designed to be “orderphillic”. As we cool the system, the crystalline phase nucleates at the orderphillic boundary and then grows until the contact line becomes pinned at the orderphillic-orderphobic boundary interface (see \autoref{fig1}d). Rotating the ring by 180°, also results in the spatial reversal of the two phases. The liquid and crystal phases are separated by a sharp stable interface ($>72$hrs) which undergoes capillary-like fluctuations, indicating the presence of a significant interfacial tension between the two phases. 
	
	The presence of a stable interface between two coexisting phases, is strongly indicative of a first-order phase transition \cite{Luu2013}. This is surprising as many experimental studies on quasi-2D granulars have found a two-step continuous transition between liquid and crystal \cite{Olafsen1998, Olafsen2005, Reis2006, Komatsu2015, Sun2016}. However, highly inelastic particles, such as ours, have previously been shown to undergo a first-order phase transition \cite{Komatsu2015}. It is important therefore to establish whether the modified transition is due to the surface dimples or the nature of the particles.  
	
\begin{figure}
		\centering
		\includegraphics{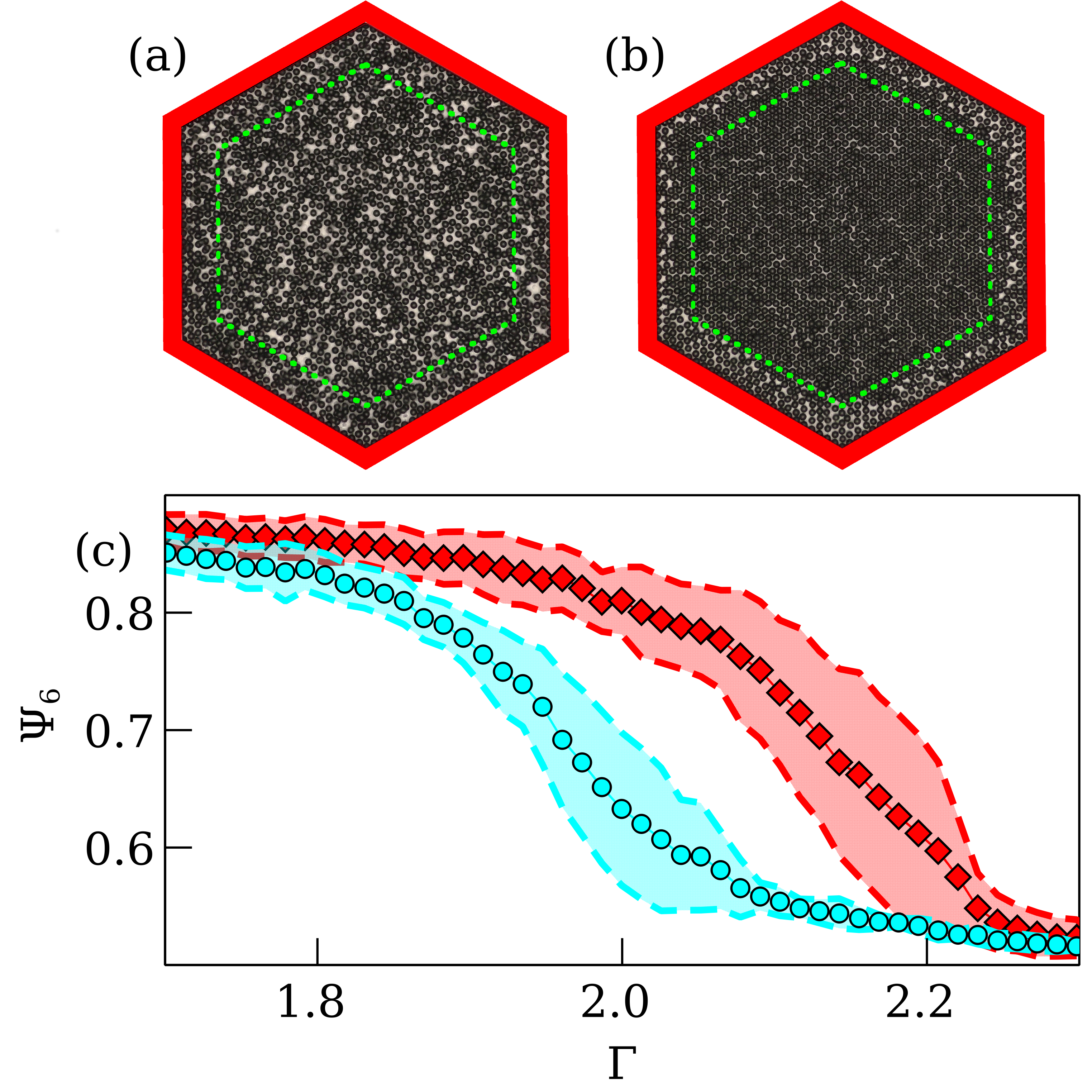}
		\caption{\textbf{Hysteresis whilst heating and cooling using a dimpled plate.} Example images (a) $\Gamma{\sim}2.4$. (b) $\Gamma{\sim}1.7$. $\psi_6$ is calculated using particles inside the green line. (c) Cooling (blue circles) and heating (red diamonds) at $\dot{\Gamma}=0.0052$s$^{-1}$. }
		\label{fig2}
	\end{figure}

	Using purely orderphobic boundaries (\autoref{fig1}a) we characterize the system’s response to cycles of cooling and heating using both the flat and dimpled surfaces. Starting with a uniform liquid state, $\phi=0.84$, $\Gamma=2.6$ (\autoref{fig2}a) the system is slowly cooled to an acceleration of $\Gamma=1.3$. With the dimpled surface, a single crystalline domain forms, separated from the boundaries due to the wetting properties (\autoref{fig2}b). The system is then slowly heated back to $\Gamma=2.6$. Each experiment is repeated 5 times. 
	
	To characterize the state of the system, we use the global orientational order parameter $\Psi_6$. This is the average value of the orientational order parameter $\psi_6^j$ for each particle j where $\psi_6^j=\frac{1}{n_j}\sum_{k=1}^{n_j}e^{6i\theta_{jk}}$.
	$n_j$ is the number of nearest neighbors determined by the Delaunay triangulation and $\theta_{jk}$ is the angle of the vector from particle j to neighboring particle k. $\Psi_6$ is calculated using those particles away from the edge (green line in \autoref{fig2}a and b).  \autoref{fig2}c shows, for the dimpled plate, how the value of $\Psi_6(\Gamma)$ changes as the system is cooled (blue circles) and heated (red diamonds) at a rate of $\dot{\Gamma}=5.2\times10^{-3}$s$^{-1}$. The standard deviation of repeat experiments is indicated by the shaded areas surrounding each curve. \autoref{fig2}c shows a pronounced hysteresis which depends on cooling rate (supplementary figure 2). The presence of hysteresis suggests the order-disorder transition is first-order \cite{Selinger2016}. Supplementary figure 3 shows the results from similar experiments using the flat plate which exhibited no hysteresis. 
	The difference in hysteresis between the flat and dimpled surface indicates that the first-order characteristics arise predominantly due to the dimples and not from particle inelasticity \cite{Komatsu2015}. We cannot, however at this point rule out the possibility that the surface tension / hysteresis on the flat plate are simply too weak to be observed. In the remainder of this paper, we seek to clarify the nature of the transitions on both surfaces.
	
	In liquids, orientational and translational order are short-range, while in crystals they are both quasi-long-range. In 2D, KTHNY theory predicts two separate transitions separated by an intermediate hexatic phase characterized by quasi-long range orientational order, but short-range translational order \cite{Nelson2002}. In contrast, in a one-step first-order phase transition the growth of both types of order occur together.

\begin{figure}
		\centering
		\includegraphics{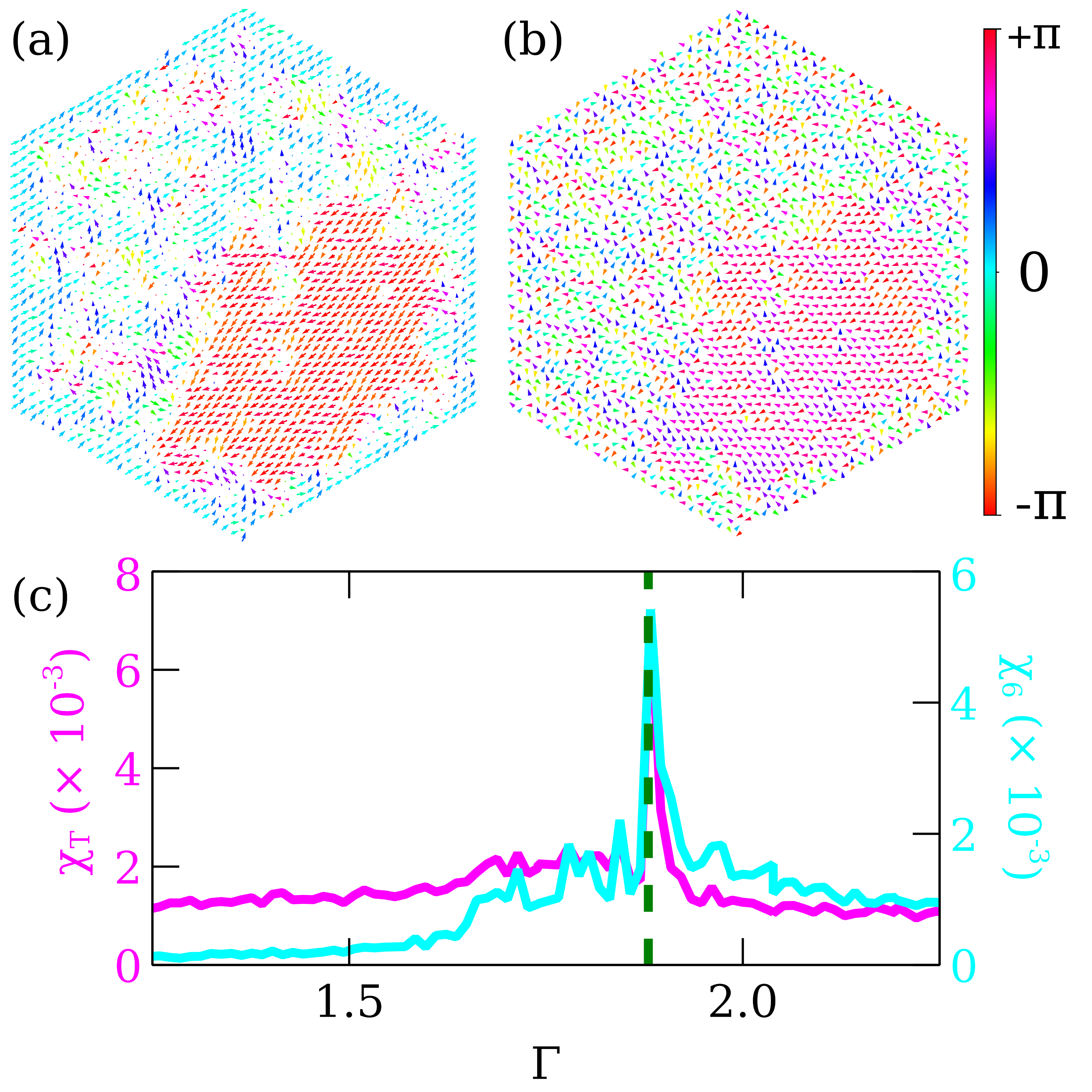}
		\caption{\textbf{Phase transition with decreasing $\Gamma$ on a dimpled plate.} Vector fields of the (a) orientational and (b) translational order parameters at $\Gamma=1.9$, cooling at $\dot{\Gamma}{\sim}2.2\times10^{-4}$s$^{-1}$. (c) The susceptibility of the translational (magenta) and orientational (cyan) order parameters.}
		\label{fig3}
	\end{figure}

\begin{figure}
		\centering
		\includegraphics{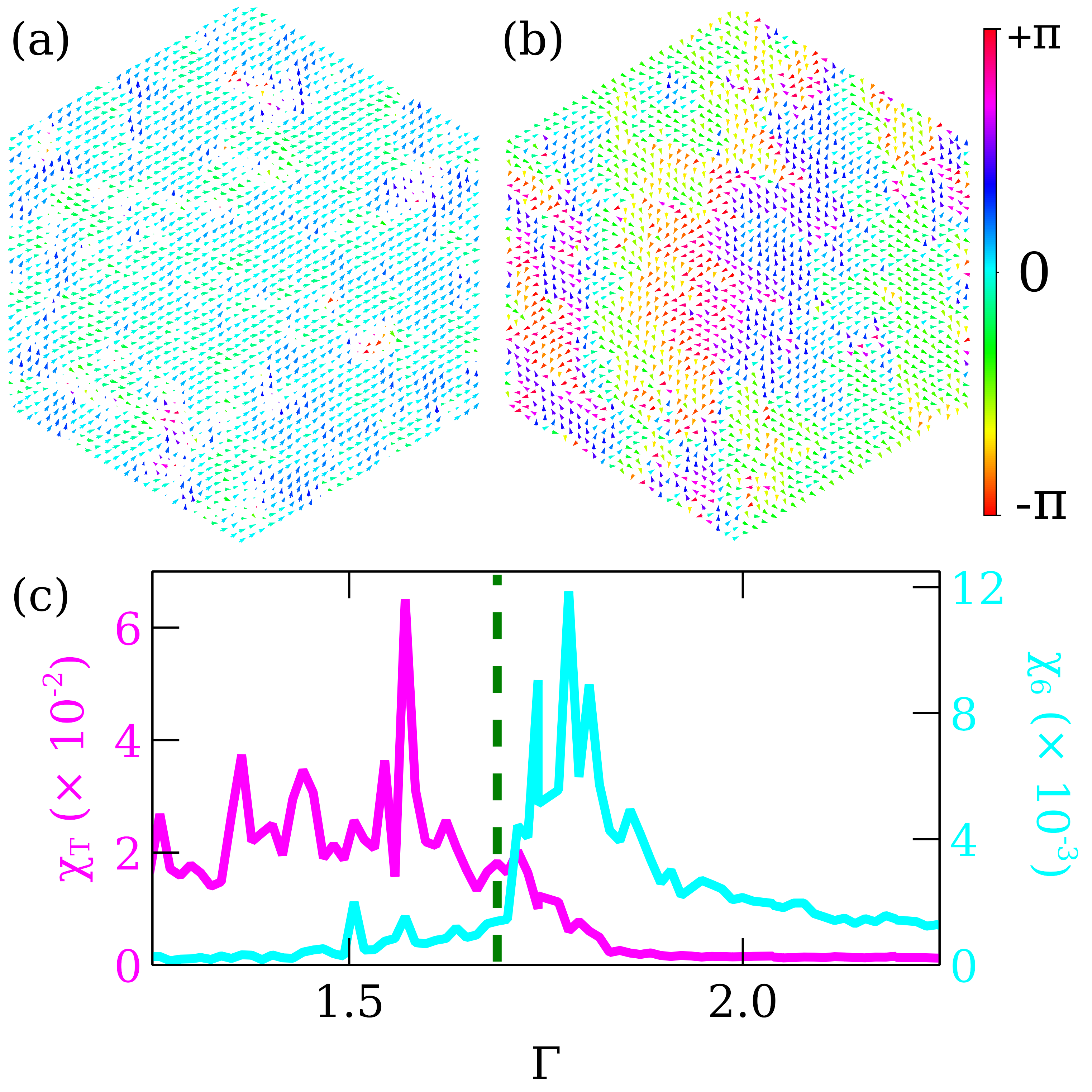}
		\caption{\textbf{Phase transition with decreasing $\Gamma$ on a flat plate.} Vector fields of the (a) orientational, (b) translational order parameters at $\Gamma=1.7$ when cooling at $\dot{\Gamma}{\sim}2.2\times10^{-4}$s$^{-1}$ on a flat plate. (c) The susceptibility of the translational (magenta) and orientational (cyan) order parameter. }
		\label{fig4}
	\end{figure}

	The local translational order of a 2D system is characterized by the order parameter $\psi_T^j=e^{i\vec{G}\cdot\vec{r}_j}$, where $\vec{r}_j$ is the position vector of particle j and $\vec{G}$ is a primary reciprocal lattice vector calculated from the Delaunay tessellation. Calculating $\psi_6^j$  and $\psi_T^j$ for each particle, we plot the corresponding vector fields, \autoref{fig3}a and b respectively. Longer arrows in the orientational order vector reflect a more ordered configuration, with the angle and color related to the orientation of the measured hexagonal order. The length of the translational order vector is always 1, while the angle and color represent the degree of translational order, such that $\pm\pi$ corresponds to particles positioned at the expected lattice points of the crystal. Plotting the vector fields allows us to visually compare changes in the spatial correlations of each order parameter for both the dimpled and flat plate (supplementary movies 2 and 3, \autoref{fig3} and \autoref{fig4}). 
	
	For the dimpled plate, at large values of the acceleration, far above the transition, both order parameters fluctuate wildly. As the acceleration is slowly reduced to $\Gamma\sim1.9$ a large area of the orientational order vector field suddenly becomes spatially correlated, indicating the formation of a crystalline nucleus (\autoref{fig3}a). At the same point in time a similar change is observed in the translational order parameter (\autoref{fig3}b). Supplementary movie 2 highlights how, at this critical acceleration, the previously uncorrelated fluctuating order parameters become both spatially and temporally synchronized in this region. It is also striking that once formed the central region of the nucleus is stable in both degree of order and orientation.

	Performing the same experiment using the flat plate (supplementary movie 3), one observes a gradual increase in the length scale of spatial fluctuations in the orientational order parameter (\autoref{fig4}a). One orientation eventually dominates the system, apparently set by the boundaries, resulting in an ordered system with small transitory patches of disorder. The spatial length scale of the translational order parameter also gradually grows as the system is cooled. However, unlike the dimpled plate, the changes of both order parameters are continuous and there are no clear spatial correlations between the two. 

	While it appears that both orientational and translational order parameters transition at a single acceleration for the dimpled plate, a more quantitative analysis is required to make confident statements concerning the flat plate. Measurements of the spatial decay of correlations in the order parameters are often used in simulations to assess where transitions occur \cite{Bernard2011, Russo2017, Li2020}. However, in experiments finite size/time effects can introduce ambiguities. Co-existing crystal and liquid phases can also result in a power-law decay that mimics a hexatic phase \cite{Han2008}. To avoid these difficulties, we measured the susceptibility of these order parameters \cite{Han2008, Sun2016}. The susceptibility is defined as  $\chi_{6/T}=\langle|\Psi_{6/T}^2|\rangle-\langle|\Psi_{6/T}|\rangle^2$, where $\Psi_{6/T}$ is the average orientational/translational order parameter of the system. \autoref{fig3}c and 4c show the translational susceptibility, $\chi_T$ (pink), and the orientational susceptibility, $\chi_6$ (blue) as the two systems are cooled.

	For the dimpled plate, the two susceptibility curves peak at the same acceleration. Together with the formation of a critical nucleus and the observed hysteresis, this confirms that the dimpled surface results in a first-order phase transition from liquid-crystal without an intervening hexatic phase. In contrast, for the flat plate (\autoref{fig4}c) the maximum fluctuations of each order parameter occur separately, indicating two separate transitions for the orientational susceptibility at $\Gamma\sim1.78$ and for the translational susceptibility at $\Gamma\sim1.57$. The presence of two distinct peaks suggests our experiment on the flat plate undergoes a two-step transition consistent with other quasi-2D granular studies \cite{Reis2006, Olafsen2005, Komatsu2015, Sun2016}. In addition, we were unable to find evidence of a finite surface tension in this system. This suggests that in our experiments on the flat plate, contrary to recent equilibrium results on colloids \cite{Thorneywork2017}, the liquid-hexatic transition is a continuous two step transition.  Importantly, in the context of this work it confirms that the first-order phase transition observed with the dimples does not merely strengthen a first-order liquid-crystal phase transition, ultimately arising from particle inelasticity, but fundamentally alters the nature of the transition. 

	In our experiments using a flat plate, highly inelastic particles ($e\sim0.1$) resulted in a two-step phase transition. This is different to the results of Komatsu et al \cite{Komatsu2015} who found that the inelasticity of their particles ($e\sim0.1$) resulted in a first-order phase transition. Subtle differences in granular experiments can radically alter observed phase behavior \cite{Castillo2012}. Indeed, Komatsu’s experiment differs from ours in having a confining lid. However, this raises the question of whether inelasticity alone is sufficient to change the order of the transition, something which warrants further investigation.

	The introduction of a dimpled surface breaks both the orientational and translational symmetry of the 2D particle fluid. However, this is only significant if the kinetic energy at which the particles undergo a liquid-solid transition is sufficiently small to be influenced by the underlying topography. The situation is like equilibrium scenarios involving noble gases on an underlying graphite lattice. There the hexatic intermediate is not observed if the liquid-solid transition occurs at thermal energies comparable to the substrate potential \cite{Aeppli1984}. If the dimples are significant, both orientational and translational ordering are affected together resulting in a single transition. One might also think that the creation of a pentagonal/heptagonal disclination, essential to the KTHNY scenario, would be much more difficult in the presence of a surface that encourages hexagonal ordering.

	In this paper we have considered the importance of order-philic/-phobic controlled wetting. We then demonstrated that a dimpled surface can result in a first-order order-disorder transition.  However, there is an important additional consequence when these phenomena occur together. In \autoref{fig1}d the use of a hybrid boundary that is orderphillic/orderphobic resulted in the symmetry of the system being broken. The order induced wetting therefore controlled the location of the resultant coexisting phases. As a system phase separates during a first-order order-disorder phase transition, order induced wetting can therefore have a long-range effect, controlling the final spatial composition of a system. Our work therefore demonstrates experimentally the importance of topography in controlling order induced interactions in 2D phase transitions. 
	
	\begin{acknowledgments}
		We thank S. Devlin and J. Pearson for technical support with equipment development. J.G.D. acknowledges a PhD studentship supported by the Royal Society. N.D.S. acknowledges a PhD studentship supported by the Engineering and Physical Sciences Research Council [Grant No. EP/M506588/1]. J.P.G acknowledges support from the EPSRC [Grant No. EP/R04421X/1]. M.I.S. gratefully acknowledges a Royal Society University Research Fellowship.
	\end{acknowledgments}
	
	\bibliographystyle{apsrev4-2}
	\bibliography{references}

\begin{thebibliography}{36}%
\makeatletter
\providecommand \@ifxundefined [1]{%
 \@ifx{#1\undefined}
}%
\providecommand \@ifnum [1]{%
 \ifnum #1\expandafter \@firstoftwo
 \else \expandafter \@secondoftwo
 \fi
}%
\providecommand \@ifx [1]{%
 \ifx #1\expandafter \@firstoftwo
 \else \expandafter \@secondoftwo
 \fi
}%
\providecommand \natexlab [1]{#1}%
\providecommand \enquote  [1]{``#1''}%
\providecommand \bibnamefont  [1]{#1}%
\providecommand \bibfnamefont [1]{#1}%
\providecommand \citenamefont [1]{#1}%
\providecommand \href@noop [0]{\@secondoftwo}%
\providecommand \href [0]{\begingroup \@sanitize@url \@href}%
\providecommand \@href[1]{\@@startlink{#1}\@@href}%
\providecommand \@@href[1]{\endgroup#1\@@endlink}%
\providecommand \@sanitize@url [0]{\catcode `\\12\catcode `\$12\catcode
  `\&12\catcode `\#12\catcode `\^12\catcode `\_12\catcode `\%12\relax}%
\providecommand \@@startlink[1]{}%
\providecommand \@@endlink[0]{}%
\providecommand \url  [0]{\begingroup\@sanitize@url \@url }%
\providecommand \@url [1]{\endgroup\@href {#1}{\urlprefix }}%
\providecommand \urlprefix  [0]{URL }%
\providecommand \Eprint [0]{\href }%
\providecommand \doibase [0]{https://doi.org/}%
\providecommand \selectlanguage [0]{\@gobble}%
\providecommand \bibinfo  [0]{\@secondoftwo}%
\providecommand \bibfield  [0]{\@secondoftwo}%
\providecommand \translation [1]{[#1]}%
\providecommand \BibitemOpen [0]{}%
\providecommand \bibitemStop [0]{}%
\providecommand \bibitemNoStop [0]{.\EOS\space}%
\providecommand \EOS [0]{\spacefactor3000\relax}%
\providecommand \BibitemShut  [1]{\csname bibitem#1\endcsname}%
\let\auto@bib@innerbib\@empty
\bibitem [{\citenamefont {Li}\ \emph {et~al.}(2021)\citenamefont {Li},
  \citenamefont {Qian},\ and\ \citenamefont {Li}}]{Li2021}%
  \BibitemOpen
  \bibfield  {author} {\bibinfo {author} {\bibfnamefont {W.}~\bibnamefont
  {Li}}, \bibinfo {author} {\bibfnamefont {X.}~\bibnamefont {Qian}},\ and\
  \bibinfo {author} {\bibfnamefont {J.}~\bibnamefont {Li}},\ }\href
  {https://doi.org/10.1038/s41578-021-00304-0} {\bibfield  {journal} {\bibinfo
  {journal} {Nat Rev Mater}\ ,\ \bibinfo {pages} {41578}} (\bibinfo {year}
  {2021})}\BibitemShut {NoStop}%
\bibitem [{\citenamefont {Wang}\ \emph {et~al.}(2018)\citenamefont {Wang},
  \citenamefont {Song}, \citenamefont {Wen}, \citenamefont {Liu}, \citenamefont
  {Wu}, \citenamefont {Dang}, \citenamefont {Hossain}, \citenamefont {Iqbal},\
  and\ \citenamefont {Xie}}]{Wang2018}%
  \BibitemOpen
  \bibfield  {author} {\bibinfo {author} {\bibfnamefont {X.}~\bibnamefont
  {Wang}}, \bibinfo {author} {\bibfnamefont {Z.}~\bibnamefont {Song}}, \bibinfo
  {author} {\bibfnamefont {W.}~\bibnamefont {Wen}}, \bibinfo {author}
  {\bibfnamefont {H.}~\bibnamefont {Liu}}, \bibinfo {author} {\bibfnamefont
  {J.}~\bibnamefont {Wu}}, \bibinfo {author} {\bibfnamefont {C.}~\bibnamefont
  {Dang}}, \bibinfo {author} {\bibfnamefont {M.}~\bibnamefont {Hossain}},
  \bibinfo {author} {\bibfnamefont {M.~A.}\ \bibnamefont {Iqbal}},\ and\
  \bibinfo {author} {\bibfnamefont {L.}~\bibnamefont {Xie}},\ }\href
  {https://doi.org/10.1002/adma.201804682} {\bibfield  {journal} {\bibinfo
  {journal} {Advanced Materials}\ }\textbf {\bibinfo {volume} {31}},\ \bibinfo
  {pages} {1804682} (\bibinfo {year} {2018})}\BibitemShut {NoStop}%
\bibitem [{\citenamefont {Zanotti}\ \emph {et~al.}(2016)\citenamefont
  {Zanotti}, \citenamefont {Judeinstein}, \citenamefont {Dalla-Bernardina},
  \citenamefont {Creff}, \citenamefont {Brubach}, \citenamefont {Roy},
  \citenamefont {Bonetti}, \citenamefont {Ollivier}, \citenamefont
  {Sakellariou},\ and\ \citenamefont {Bellissent-Funel}}]{Zanotti2016}%
  \BibitemOpen
  \bibfield  {author} {\bibinfo {author} {\bibfnamefont {J.-M.}\ \bibnamefont
  {Zanotti}}, \bibinfo {author} {\bibfnamefont {P.}~\bibnamefont
  {Judeinstein}}, \bibinfo {author} {\bibfnamefont {S.}~\bibnamefont
  {Dalla-Bernardina}}, \bibinfo {author} {\bibfnamefont {G.}~\bibnamefont
  {Creff}}, \bibinfo {author} {\bibfnamefont {J.-B.}\ \bibnamefont {Brubach}},
  \bibinfo {author} {\bibfnamefont {P.}~\bibnamefont {Roy}}, \bibinfo {author}
  {\bibfnamefont {M.}~\bibnamefont {Bonetti}}, \bibinfo {author} {\bibfnamefont
  {J.}~\bibnamefont {Ollivier}}, \bibinfo {author} {\bibfnamefont
  {D.}~\bibnamefont {Sakellariou}},\ and\ \bibinfo {author} {\bibfnamefont
  {M.-C.}\ \bibnamefont {Bellissent-Funel}},\ }\href
  {https://doi.org/10.1038/srep25938} {\bibfield  {journal} {\bibinfo
  {journal} {Sci Rep}\ }\textbf {\bibinfo {volume} {6}},\ \bibinfo {pages}
  {25938} (\bibinfo {year} {2016})}\BibitemShut {NoStop}%
\bibitem [{\citenamefont {Durand}\ and\ \citenamefont
  {Heu}(2019)}]{Durand2019}%
  \BibitemOpen
  \bibfield  {author} {\bibinfo {author} {\bibfnamefont {M.}~\bibnamefont
  {Durand}}\ and\ \bibinfo {author} {\bibfnamefont {J.}~\bibnamefont {Heu}},\
  }\href {https://doi.org/10.1103/PhysRevLett.123.188001} {\bibfield  {journal}
  {\bibinfo  {journal} {Phys. Rev. Lett.}\ }\textbf {\bibinfo {volume} {123}},\
  \bibinfo {pages} {188001} (\bibinfo {year} {2019})}\BibitemShut {NoStop}%
\bibitem [{\citenamefont {Bernard}\ and\ \citenamefont
  {Krauth}(2011)}]{Bernard2011}%
  \BibitemOpen
  \bibfield  {author} {\bibinfo {author} {\bibfnamefont {E.~P.}\ \bibnamefont
  {Bernard}}\ and\ \bibinfo {author} {\bibfnamefont {W.}~\bibnamefont
  {Krauth}},\ }\href {https://doi.org/10.1103/PhysRevLett.107.155704}
  {\bibfield  {journal} {\bibinfo  {journal} {Phys. Rev. Lett.}\ }\textbf
  {\bibinfo {volume} {107}},\ \bibinfo {pages} {155704} (\bibinfo {year}
  {2011})}\BibitemShut {NoStop}%
\bibitem [{\citenamefont {Halperin}\ and\ \citenamefont
  {Nelson}(1978)}]{Halperin1978}%
  \BibitemOpen
  \bibfield  {author} {\bibinfo {author} {\bibfnamefont {B.}~\bibnamefont
  {Halperin}}\ and\ \bibinfo {author} {\bibfnamefont {D.}~\bibnamefont
  {Nelson}},\ }\href {https://doi.org/10.1103/physrevlett.41.121} {\bibfield
  {journal} {\bibinfo  {journal} {Phys. Rev. Lett.}\ }\textbf {\bibinfo
  {volume} {41}},\ \bibinfo {pages} {121} (\bibinfo {year} {1978})}\BibitemShut
  {NoStop}%
\bibitem [{\citenamefont {Engel}\ \emph {et~al.}(2013)\citenamefont {Engel},
  \citenamefont {Anderson}, \citenamefont {Glotzer}, \citenamefont {Isobe},
  \citenamefont {Bernard},\ and\ \citenamefont {Krauth}}]{Engel2013}%
  \BibitemOpen
  \bibfield  {author} {\bibinfo {author} {\bibfnamefont {M.}~\bibnamefont
  {Engel}}, \bibinfo {author} {\bibfnamefont {J.~A.}\ \bibnamefont {Anderson}},
  \bibinfo {author} {\bibfnamefont {S.~C.}\ \bibnamefont {Glotzer}}, \bibinfo
  {author} {\bibfnamefont {M.}~\bibnamefont {Isobe}}, \bibinfo {author}
  {\bibfnamefont {E.~P.}\ \bibnamefont {Bernard}},\ and\ \bibinfo {author}
  {\bibfnamefont {W.}~\bibnamefont {Krauth}},\ }\href
  {https://doi.org/10.1103/PhysRevE.87.042134} {\bibfield  {journal} {\bibinfo
  {journal} {Phys. Rev. E}\ }\textbf {\bibinfo {volume} {87}},\ \bibinfo
  {pages} {042134} (\bibinfo {year} {2013})}\BibitemShut {NoStop}%
\bibitem [{\citenamefont {Kapfer}\ and\ \citenamefont
  {Krauth}(2015)}]{Kapfer2015}%
  \BibitemOpen
  \bibfield  {author} {\bibinfo {author} {\bibfnamefont {S.~C.}\ \bibnamefont
  {Kapfer}}\ and\ \bibinfo {author} {\bibfnamefont {W.}~\bibnamefont
  {Krauth}},\ }\href {https://doi.org/10.1103/physrevlett.114.035702}
  {\bibfield  {journal} {\bibinfo  {journal} {Phys. Rev. Lett.}\ }\textbf
  {\bibinfo {volume} {114}},\ \bibinfo {pages} {035702} (\bibinfo {year}
  {2015})}\BibitemShut {NoStop}%
\bibitem [{\citenamefont {Russo}\ and\ \citenamefont
  {Wilding}(2017)}]{Russo2017}%
  \BibitemOpen
  \bibfield  {author} {\bibinfo {author} {\bibfnamefont {J.}~\bibnamefont
  {Russo}}\ and\ \bibinfo {author} {\bibfnamefont {N.~B.}\ \bibnamefont
  {Wilding}},\ }\href {https://doi.org/10.1103/PhysRevLett.119.115702}
  {\bibfield  {journal} {\bibinfo  {journal} {Phys. Rev. Lett.}\ }\textbf
  {\bibinfo {volume} {119}},\ \bibinfo {pages} {115702} (\bibinfo {year}
  {2017})}\BibitemShut {NoStop}%
\bibitem [{\citenamefont {Anderson}\ \emph {et~al.}(2017)\citenamefont
  {Anderson}, \citenamefont {Antonaglia}, \citenamefont {Millan}, \citenamefont
  {Engel},\ and\ \citenamefont {Glotzer}}]{Anderson2017}%
  \BibitemOpen
  \bibfield  {author} {\bibinfo {author} {\bibfnamefont {J.~A.}\ \bibnamefont
  {Anderson}}, \bibinfo {author} {\bibfnamefont {J.}~\bibnamefont
  {Antonaglia}}, \bibinfo {author} {\bibfnamefont {J.~A.}\ \bibnamefont
  {Millan}}, \bibinfo {author} {\bibfnamefont {M.}~\bibnamefont {Engel}},\ and\
  \bibinfo {author} {\bibfnamefont {S.~C.}\ \bibnamefont {Glotzer}},\ }\href
  {https://doi.org/10.1103/physrevx.7.021001} {\bibfield  {journal} {\bibinfo
  {journal} {Phys. Rev. X}\ }\textbf {\bibinfo {volume} {7}},\ \bibinfo {pages}
  {021001} (\bibinfo {year} {2017})}\BibitemShut {NoStop}%
\bibitem [{\citenamefont {Li}\ and\ \citenamefont {Ciamarra}(2020)}]{Li2020}%
  \BibitemOpen
  \bibfield  {author} {\bibinfo {author} {\bibfnamefont {Y.-W.}\ \bibnamefont
  {Li}}\ and\ \bibinfo {author} {\bibfnamefont {M.~P.}\ \bibnamefont
  {Ciamarra}},\ }\href {https://doi.org/10.1103/PhysRevLett.124.218002}
  {\bibfield  {journal} {\bibinfo  {journal} {Phys. Rev. Lett.}\ }\textbf
  {\bibinfo {volume} {124}},\ \bibinfo {pages} {218002} (\bibinfo {year}
  {2020})}\BibitemShut {NoStop}%
\bibitem [{\citenamefont {Briand}\ and\ \citenamefont
  {Dauchot}(2016)}]{Briand2016}%
  \BibitemOpen
  \bibfield  {author} {\bibinfo {author} {\bibfnamefont {G.}~\bibnamefont
  {Briand}}\ and\ \bibinfo {author} {\bibfnamefont {O.}~\bibnamefont
  {Dauchot}},\ }\href {https://doi.org/10.1103/PhysRevLett.117.098004}
  {\bibfield  {journal} {\bibinfo  {journal} {Phys. Rev. Lett.}\ }\textbf
  {\bibinfo {volume} {117}},\ \bibinfo {pages} {098004} (\bibinfo {year}
  {2016})}\BibitemShut {NoStop}%
\bibitem [{\citenamefont {Thorneywork}\ \emph {et~al.}(2017)\citenamefont
  {Thorneywork}, \citenamefont {Abbott}, \citenamefont {Aarts},\ and\
  \citenamefont {Dullens}}]{Thorneywork2017}%
  \BibitemOpen
  \bibfield  {author} {\bibinfo {author} {\bibfnamefont {A.~L.}\ \bibnamefont
  {Thorneywork}}, \bibinfo {author} {\bibfnamefont {J.~L.}\ \bibnamefont
  {Abbott}}, \bibinfo {author} {\bibfnamefont {D.~G.}\ \bibnamefont {Aarts}},\
  and\ \bibinfo {author} {\bibfnamefont {R.~P.}\ \bibnamefont {Dullens}},\
  }\href {https://doi.org/10.1103/physrevlett.118.158001} {\bibfield  {journal}
  {\bibinfo  {journal} {Phys. Rev. Lett.}\ }\textbf {\bibinfo {volume} {118}},\
  \bibinfo {pages} {158001} (\bibinfo {year} {2017})}\BibitemShut {NoStop}%
\bibitem [{\citenamefont {Li}\ \emph {et~al.}(2019)\citenamefont {Li},
  \citenamefont {Xiao}, \citenamefont {Wang}, \citenamefont {Wen},\ and\
  \citenamefont {Wang}}]{Li2019}%
  \BibitemOpen
  \bibfield  {author} {\bibinfo {author} {\bibfnamefont {B.}~\bibnamefont
  {Li}}, \bibinfo {author} {\bibfnamefont {X.}~\bibnamefont {Xiao}}, \bibinfo
  {author} {\bibfnamefont {S.}~\bibnamefont {Wang}}, \bibinfo {author}
  {\bibfnamefont {W.}~\bibnamefont {Wen}},\ and\ \bibinfo {author}
  {\bibfnamefont {Z.}~\bibnamefont {Wang}},\ }\href
  {https://doi.org/10.1103/PhysRevX.9.031032} {\bibfield  {journal} {\bibinfo
  {journal} {Phys. Rev. X}\ }\textbf {\bibinfo {volume} {9}},\ \bibinfo {pages}
  {031032} (\bibinfo {year} {2019})}\BibitemShut {NoStop}%
\bibitem [{\citenamefont {Olafsen}\ and\ \citenamefont
  {Urbach}(1998)}]{Olafsen1998}%
  \BibitemOpen
  \bibfield  {author} {\bibinfo {author} {\bibfnamefont {J.~S.}\ \bibnamefont
  {Olafsen}}\ and\ \bibinfo {author} {\bibfnamefont {J.~S.}\ \bibnamefont
  {Urbach}},\ }\href {https://doi.org/10.1103/PhysRevLett.81.4369} {\bibfield
  {journal} {\bibinfo  {journal} {Phys. Rev. Lett.}\ }\textbf {\bibinfo
  {volume} {81}},\ \bibinfo {pages} {4369} (\bibinfo {year}
  {1998})}\BibitemShut {NoStop}%
\bibitem [{\citenamefont {Olafsen}\ and\ \citenamefont
  {Urbach}(2005)}]{Olafsen2005}%
  \BibitemOpen
  \bibfield  {author} {\bibinfo {author} {\bibfnamefont {J.~S.}\ \bibnamefont
  {Olafsen}}\ and\ \bibinfo {author} {\bibfnamefont {J.~S.}\ \bibnamefont
  {Urbach}},\ }\href {https://doi.org/10.1103/physrevlett.95.098002} {\bibfield
   {journal} {\bibinfo  {journal} {Phys. Rev. Lett.}\ }\textbf {\bibinfo
  {volume} {95}},\ \bibinfo {pages} {098002} (\bibinfo {year}
  {2005})}\BibitemShut {NoStop}%
\bibitem [{\citenamefont {Reis}\ \emph {et~al.}(2006)\citenamefont {Reis},
  \citenamefont {Ingale},\ and\ \citenamefont {Shattuck}}]{Reis2006}%
  \BibitemOpen
  \bibfield  {author} {\bibinfo {author} {\bibfnamefont {P.~M.}\ \bibnamefont
  {Reis}}, \bibinfo {author} {\bibfnamefont {R.~A.}\ \bibnamefont {Ingale}},\
  and\ \bibinfo {author} {\bibfnamefont {M.~D.}\ \bibnamefont {Shattuck}},\
  }\href {https://doi.org/10.1103/PhysRevLett.96.258001} {\bibfield  {journal}
  {\bibinfo  {journal} {Phys. Rev. Lett.}\ }\textbf {\bibinfo {volume} {96}},\
  \bibinfo {pages} {258001} (\bibinfo {year} {2006})}\BibitemShut {NoStop}%
\bibitem [{\citenamefont {Komatsu}\ and\ \citenamefont
  {Tanaka}(2015)}]{Komatsu2015}%
  \BibitemOpen
  \bibfield  {author} {\bibinfo {author} {\bibfnamefont {Y.}~\bibnamefont
  {Komatsu}}\ and\ \bibinfo {author} {\bibfnamefont {H.}~\bibnamefont
  {Tanaka}},\ }\href {https://doi.org/10.1103/PhysRevX.5.031025} {\bibfield
  {journal} {\bibinfo  {journal} {Phys. Rev. X}\ }\textbf {\bibinfo {volume}
  {5}},\ \bibinfo {pages} {031025} (\bibinfo {year} {2015})}\BibitemShut
  {NoStop}%
\bibitem [{\citenamefont {Sun}\ \emph {et~al.}(2016)\citenamefont {Sun},
  \citenamefont {Li}, \citenamefont {Ma},\ and\ \citenamefont
  {Zhang}}]{Sun2016}%
  \BibitemOpen
  \bibfield  {author} {\bibinfo {author} {\bibfnamefont {X.}~\bibnamefont
  {Sun}}, \bibinfo {author} {\bibfnamefont {Y.}~\bibnamefont {Li}}, \bibinfo
  {author} {\bibfnamefont {Y.}~\bibnamefont {Ma}},\ and\ \bibinfo {author}
  {\bibfnamefont {Z.}~\bibnamefont {Zhang}},\ }\href
  {https://doi.org/10.1038/srep24056} {\bibfield  {journal} {\bibinfo
  {journal} {Sci Rep}\ }\textbf {\bibinfo {volume} {6}},\ \bibinfo {pages}
  {24056} (\bibinfo {year} {2016})}\BibitemShut {NoStop}%
\bibitem [{\citenamefont {Nelson}(2002)}]{Nelson2002}%
  \BibitemOpen
  \bibfield  {author} {\bibinfo {author} {\bibfnamefont {D.~R.}\ \bibnamefont
  {Nelson}},\ }\href
  {https://www.ebook.de/de/product/2784975/david_r_nelson_defects_and_geometry_in_condensed_matter_physics.html}
  {\emph {\bibinfo {title} {Defects and Geometry in Condensed Matter
  Physics}}}\ (\bibinfo  {publisher} {Cambridge University Press},\ \bibinfo
  {year} {2002})\BibitemShut {NoStop}%
\bibitem [{\citenamefont {Katira}\ \emph {et~al.}(2016)\citenamefont {Katira},
  \citenamefont {Mandadapu}, \citenamefont {Vaikuntanathan}, \citenamefont
  {Smit},\ and\ \citenamefont {Chandler}}]{Katira2016}%
  \BibitemOpen
  \bibfield  {author} {\bibinfo {author} {\bibfnamefont {S.}~\bibnamefont
  {Katira}}, \bibinfo {author} {\bibfnamefont {K.~K.}\ \bibnamefont
  {Mandadapu}}, \bibinfo {author} {\bibfnamefont {S.}~\bibnamefont
  {Vaikuntanathan}}, \bibinfo {author} {\bibfnamefont {B.}~\bibnamefont
  {Smit}},\ and\ \bibinfo {author} {\bibfnamefont {D.}~\bibnamefont
  {Chandler}},\ }\href {https://doi.org/10.7554/elife.13150} {\bibfield
  {journal} {\bibinfo  {journal} {{eLife}}\ }\textbf {\bibinfo {volume} {5}},\
  \bibinfo {pages} {13150} (\bibinfo {year} {2016})}\BibitemShut {NoStop}%
\bibitem [{\citenamefont {Katira}\ \emph {et~al.}(2018)\citenamefont {Katira},
  \citenamefont {Garrahan},\ and\ \citenamefont {Mandadapu}}]{Katira2018}%
  \BibitemOpen
  \bibfield  {author} {\bibinfo {author} {\bibfnamefont {S.}~\bibnamefont
  {Katira}}, \bibinfo {author} {\bibfnamefont {J.~P.}\ \bibnamefont
  {Garrahan}},\ and\ \bibinfo {author} {\bibfnamefont {K.~K.}\ \bibnamefont
  {Mandadapu}},\ }\href {https://doi.org/10.1103/PhysRevLett.120.260602}
  {\bibfield  {journal} {\bibinfo  {journal} {Phys. Rev. Lett.}\ }\textbf
  {\bibinfo {volume} {120}},\ \bibinfo {pages} {260602} (\bibinfo {year}
  {2018})}\BibitemShut {NoStop}%
\bibitem [{\citenamefont {Birgeneau}\ \emph {et~al.}(1981)\citenamefont
  {Birgeneau}, \citenamefont {Brown}, \citenamefont {Horn},\ and\ \citenamefont
  {Moncton}}]{Birgeneau1981}%
  \BibitemOpen
  \bibfield  {author} {\bibinfo {author} {\bibfnamefont {R.~J.}\ \bibnamefont
  {Birgeneau}}, \bibinfo {author} {\bibfnamefont {G.~S.}\ \bibnamefont
  {Brown}}, \bibinfo {author} {\bibfnamefont {P.~M.}\ \bibnamefont {Horn}},\
  and\ \bibinfo {author} {\bibfnamefont {D.~E.}\ \bibnamefont {Moncton}},\
  }\href {https://doi.org/10.1088/0022-3719/14/3/001} {\bibfield  {journal}
  {\bibinfo  {journal} {J. phys., C, Solid state phys}\ }\textbf {\bibinfo
  {volume} {14}},\ \bibinfo {pages} {L49} (\bibinfo {year} {1981})}\BibitemShut
  {NoStop}%
\bibitem [{\citenamefont {Strandburg}(1988)}]{Strandburg1988}%
  \BibitemOpen
  \bibfield  {author} {\bibinfo {author} {\bibfnamefont {K.~J.}\ \bibnamefont
  {Strandburg}},\ }\href {https://doi.org/10.1103/revmodphys.60.161} {\bibfield
   {journal} {\bibinfo  {journal} {Rev. Mod. Phys.}\ }\textbf {\bibinfo
  {volume} {60}},\ \bibinfo {pages} {161} (\bibinfo {year} {1988})}\BibitemShut
  {NoStop}%
\bibitem [{\citenamefont {Ganapathy}\ \emph {et~al.}(2010)\citenamefont
  {Ganapathy}, \citenamefont {Buckley}, \citenamefont {Gerbode},\ and\
  \citenamefont {Cohen}}]{Ganapathy2010}%
  \BibitemOpen
  \bibfield  {author} {\bibinfo {author} {\bibfnamefont {R.}~\bibnamefont
  {Ganapathy}}, \bibinfo {author} {\bibfnamefont {M.~R.}\ \bibnamefont
  {Buckley}}, \bibinfo {author} {\bibfnamefont {S.~J.}\ \bibnamefont
  {Gerbode}},\ and\ \bibinfo {author} {\bibfnamefont {I.}~\bibnamefont
  {Cohen}},\ }\href {https://doi.org/10.1126/science.1179947} {\bibfield
  {journal} {\bibinfo  {journal} {Science}\ }\textbf {\bibinfo {volume}
  {327}},\ \bibinfo {pages} {445} (\bibinfo {year} {2010})}\BibitemShut
  {NoStop}%
\bibitem [{\citenamefont {Mishra}\ \emph {et~al.}(2016)\citenamefont {Mishra},
  \citenamefont {Sood},\ and\ \citenamefont {Ganapathy}}]{Mishra2016}%
  \BibitemOpen
  \bibfield  {author} {\bibinfo {author} {\bibfnamefont {C.~K.}\ \bibnamefont
  {Mishra}}, \bibinfo {author} {\bibfnamefont {A.~K.}\ \bibnamefont {Sood}},\
  and\ \bibinfo {author} {\bibfnamefont {R.}~\bibnamefont {Ganapathy}},\ }\href
  {https://doi.org/10.1073/pnas.1608568113} {\bibfield  {journal} {\bibinfo
  {journal} {PNAS}\ }\textbf {\bibinfo {volume} {113}},\ \bibinfo {pages}
  {12094} (\bibinfo {year} {2016})}\BibitemShut {NoStop}%
\bibitem [{\citenamefont {Qi}\ and\ \citenamefont {Dijkstra}(2015)}]{Qi2015}%
  \BibitemOpen
  \bibfield  {author} {\bibinfo {author} {\bibfnamefont {W.}~\bibnamefont
  {Qi}}\ and\ \bibinfo {author} {\bibfnamefont {M.}~\bibnamefont {Dijkstra}},\
  }\href {https://doi.org/10.1039/C4SM02876G} {\bibfield  {journal} {\bibinfo
  {journal} {Soft Matter}\ }\textbf {\bibinfo {volume} {11}},\ \bibinfo {pages}
  {2852} (\bibinfo {year} {2015})}\BibitemShut {NoStop}%
\bibitem [{\citenamefont {Clewett}\ \emph {et~al.}(2019)\citenamefont
  {Clewett}, \citenamefont {Bowley},\ and\ \citenamefont
  {Swift}}]{Clewett2019}%
  \BibitemOpen
  \bibfield  {author} {\bibinfo {author} {\bibfnamefont {J.~P.}\ \bibnamefont
  {Clewett}}, \bibinfo {author} {\bibfnamefont {R.}~\bibnamefont {Bowley}},\
  and\ \bibinfo {author} {\bibfnamefont {M.~R.}\ \bibnamefont {Swift}},\ }\href
  {https://doi.org/10.1103/PhysRevLett.123.118001} {\bibfield  {journal}
  {\bibinfo  {journal} {Phys. Rev. Lett.}\ }\textbf {\bibinfo {volume} {123}},\
  \bibinfo {pages} {118001} (\bibinfo {year} {2019})}\BibitemShut {NoStop}%
\bibitem [{\citenamefont {Smith}\ and\ \citenamefont
  {Smith}(2017)}]{Smith2017}%
  \BibitemOpen
  \bibfield  {author} {\bibinfo {author} {\bibfnamefont {N.~D.}\ \bibnamefont
  {Smith}}\ and\ \bibinfo {author} {\bibfnamefont {M.~I.}\ \bibnamefont
  {Smith}},\ }\href {https://doi.org/10.1103/physreve.96.062910} {\bibfield
  {journal} {\bibinfo  {journal} {Phys. Rev. E}\ }\textbf {\bibinfo {volume}
  {96}},\ \bibinfo {pages} {062910} (\bibinfo {year} {2017})}\BibitemShut
  {NoStop}%
\bibitem [{\citenamefont {Galajda}\ \emph {et~al.}(2007)\citenamefont
  {Galajda}, \citenamefont {Keymer}, \citenamefont {Chaikin},\ and\
  \citenamefont {Austin}}]{Galajda2007}%
  \BibitemOpen
  \bibfield  {author} {\bibinfo {author} {\bibfnamefont {P.}~\bibnamefont
  {Galajda}}, \bibinfo {author} {\bibfnamefont {J.}~\bibnamefont {Keymer}},
  \bibinfo {author} {\bibfnamefont {P.}~\bibnamefont {Chaikin}},\ and\ \bibinfo
  {author} {\bibfnamefont {R.}~\bibnamefont {Austin}},\ }\href
  {https://doi.org/10.1128/JB.01033-07} {\bibfield  {journal} {\bibinfo
  {journal} {J Bacteriol}\ }\textbf {\bibinfo {volume} {189}},\ \bibinfo
  {pages} {8704} (\bibinfo {year} {2007})}\BibitemShut {NoStop}%
\bibitem [{\citenamefont {Reichhardt}\ and\ \citenamefont
  {Reichhardt}(2021)}]{Reichhardt2021}%
  \BibitemOpen
  \bibfield  {author} {\bibinfo {author} {\bibfnamefont {C.}~\bibnamefont
  {Reichhardt}}\ and\ \bibinfo {author} {\bibfnamefont {C.~J.~O.}\ \bibnamefont
  {Reichhardt}},\ }\href {https://doi.org/10.1103/PhysRevE.103.022602}
  {\bibfield  {journal} {\bibinfo  {journal} {Phys. Rev. E}\ }\textbf {\bibinfo
  {volume} {103}},\ \bibinfo {pages} {022602} (\bibinfo {year}
  {2021})}\BibitemShut {NoStop}%
\bibitem [{\citenamefont {Luu}\ \emph {et~al.}(2013)\citenamefont {Luu},
  \citenamefont {Castillo}, \citenamefont {Mujica},\ and\ \citenamefont
  {Soto}}]{Luu2013}%
  \BibitemOpen
  \bibfield  {author} {\bibinfo {author} {\bibfnamefont {L.-H.}\ \bibnamefont
  {Luu}}, \bibinfo {author} {\bibfnamefont {G.}~\bibnamefont {Castillo}},
  \bibinfo {author} {\bibfnamefont {N.}~\bibnamefont {Mujica}},\ and\ \bibinfo
  {author} {\bibfnamefont {R.}~\bibnamefont {Soto}},\ }\href
  {https://doi.org/10.1103/PhysRevE.87.040202} {\bibfield  {journal} {\bibinfo
  {journal} {Phys. Rev. E}\ }\textbf {\bibinfo {volume} {87}},\ \bibinfo
  {pages} {040202} (\bibinfo {year} {2013})}\BibitemShut {NoStop}%
\bibitem [{\citenamefont {Selinger}(2016)}]{Selinger2016}%
  \BibitemOpen
  \bibfield  {author} {\bibinfo {author} {\bibfnamefont {J.~V.}\ \bibnamefont
  {Selinger}},\ }\href {https://doi.org/10.1007/978-3-319-21054-4} {\emph
  {\bibinfo {title} {Introduction to the Theory of Soft Matter}}}\ (\bibinfo
  {publisher} {Springer International Publishing},\ \bibinfo {year}
  {2016})\BibitemShut {NoStop}%
\bibitem [{\citenamefont {Han}\ \emph {et~al.}(2008)\citenamefont {Han},
  \citenamefont {Ha}, \citenamefont {Alsayed},\ and\ \citenamefont
  {Yodh}}]{Han2008}%
  \BibitemOpen
  \bibfield  {author} {\bibinfo {author} {\bibfnamefont {Y.}~\bibnamefont
  {Han}}, \bibinfo {author} {\bibfnamefont {N.~Y.}\ \bibnamefont {Ha}},
  \bibinfo {author} {\bibfnamefont {A.~M.}\ \bibnamefont {Alsayed}},\ and\
  \bibinfo {author} {\bibfnamefont {A.~G.}\ \bibnamefont {Yodh}},\ }\href
  {https://doi.org/10.1103/PhysRevE.77.041406} {\bibfield  {journal} {\bibinfo
  {journal} {Phys. Rev. E}\ }\textbf {\bibinfo {volume} {77}},\ \bibinfo
  {pages} {041406} (\bibinfo {year} {2008})}\BibitemShut {NoStop}%
\bibitem [{\citenamefont {Castillo}\ \emph {et~al.}(2012)\citenamefont
  {Castillo}, \citenamefont {Mujica},\ and\ \citenamefont
  {Soto}}]{Castillo2012}%
  \BibitemOpen
  \bibfield  {author} {\bibinfo {author} {\bibfnamefont {G.}~\bibnamefont
  {Castillo}}, \bibinfo {author} {\bibfnamefont {N.}~\bibnamefont {Mujica}},\
  and\ \bibinfo {author} {\bibfnamefont {R.}~\bibnamefont {Soto}},\ }\href
  {https://doi.org/10.1103/PhysRevLett.109.095701} {\bibfield  {journal}
  {\bibinfo  {journal} {Phys. Rev. Lett.}\ }\textbf {\bibinfo {volume} {109}},\
  \bibinfo {pages} {095701} (\bibinfo {year} {2012})}\BibitemShut {NoStop}%
\bibitem [{\citenamefont {Aeppli}\ and\ \citenamefont
  {Bruinsma}(1984)}]{Aeppli1984}%
  \BibitemOpen
  \bibfield  {author} {\bibinfo {author} {\bibfnamefont {G.}~\bibnamefont
  {Aeppli}}\ and\ \bibinfo {author} {\bibfnamefont {R.}~\bibnamefont
  {Bruinsma}},\ }\href {https://doi.org/10.1103/PhysRevLett.53.2133} {\bibfield
   {journal} {\bibinfo  {journal} {Phys. Rev. Lett.}\ }\textbf {\bibinfo
  {volume} {53}},\ \bibinfo {pages} {2133} (\bibinfo {year}
  {1984})}\BibitemShut {NoStop}%
\end{thebibliography}%
	
\end{document}